%Paper: 9203070
%From: papa@qmchep.dnet.nasa.gov
%Date: Wed, 25 Mar 92 11:34:49 -0500

\input phyzzx

\Pubnum={QMW 92/04}
\pubtype={}
\date={March 1992}

\titlepage

\title{Finiteness and
anomalies in $(4,0)$ supersymmetric sigma models}

\author{ P.S. Howe }
\address {Department of Mathematics, King's College, London, UK.}
\andauthor {G. Papadopoulos}
\address{Deptartment of Physics, Queen Mary and Westfield College, London, UK.}

\abstract{Power-counting arguments based on extended superfields
have been used to argue
that two-dimensional supersymmetric sigma models with
(4,0) supersymmetry are finite. This result is confirmed up to
three loop order in perturbation theory by an explicit calculation
using (1,0) superfields. In particular, it is shown that the finite
counterterms which must be introduced into the theory in order to
maintain (4,0) supersymmetry are precisely the terms that are required
to establish ultra-violet finiteness.}

\endpage

\pagenumber=2

%%%%%%%%%%%%%%%%%%%%%%%%%%%%%%%% MACROS %%%%%%%%%%%%%%%%%%%%%%%%%%%%%%%%%%%%%%%

\def\fff{\vrule width0.5pt height5pt depth1pt}
\def\pp{{{ =\hskip-3.75pt{\fff}}\hskip3.75pt }}

\def\gij{g_{ij}}

\def\bij{b_{ij}}
\def\ffi{\phi\sp{i}}
\def\ffj{\phi\sp{j}}

\def\del{\partial}

\def\th{\theta}

\def\ap{\alpha'}
\def\app{(\alpha')}

%%%%%%%%%%%%%%%%%%%%%%%%%%%%% REFERENCES %%%%%%%%%%%%%%%%%%%%%%%%%%%%%%%%%%%%%%

\REF\hks {P.S. Howe and K.S. Stelle, Int. J. Mod. Phys. \underbar{
A4} (1989) 1871.}

\REF\agf {L. Alvarez-Gaum\'e and D.Z. Freedman, Comm. Math. Phys.
\underbar{80} (1981) 443.}

\REF\cmha {C.M. Hull, Nucl. Phys. \underbar{B260} (1985) 182.}

\REF\gios {A. Galperin, E. Ivanov, V. Ogievetsky and E. Sokatchev,
Class. Quant. Grav. \underbar{2} (1985) 617.}

 \REF\hpa  {P.S. Howe and G. Papadopoulos, Nucl. Phys. \underbar {B289}
 (1987), 264.}

\REF\hpb  {P.S. Howe and G. Papadopoulos, Class. Quantum Grav. \underbar {5}
(1988), 1647.}

\REF\kss {E. Sokatchev and K.S. Stelle, Class. Quant. Grav. \underbar{4}
(1987) 501.}

\REF\bp  {C. Becchi and O. Piguet, Nucl. Phys. \underbar
{B347} (1990), 596.}

\REF\s{A. Strominger, Nucl.Phys. \underbar{B343} (1990) 167.}

\REF\hs{J. Harvey and A. Strominger, Phys. Rev. Lett. \underbar{66}
(1991) 549.}

\REF\chsa {C. G. Callan, Jr., J. A. Harvey, and A. Strominger,
Nucl. Phys. \underbar{B359} (1991) 611.}

\REF\chsb {C. G. Callan, Jr., J. A. Harvey, and A. Strominger,
{\sl Worldbrane actions for string solitons}, PUTP-1244, EFI-91-12
(1991).}

\REF\chsc {C. G. Callan, Jr., J. A. Harvey, and A. Strominger,
{\sl Supersymmetric String Solitons}, lectures given at the 1991
Trieste Sping School, EFI-91-66 (1991).}

\REF\dl {M.J. Duff and J. Lu, Nucl. Phys. \underbar{B354} (1991)
141.}

\REF\pt {I. Pesando and A. Tollst\'en, CERN-TH 6286/91, DFTT 42/91,
(1991).}
\REF\eoty {T. Eguchi, H. Ooguri, A. Taormina and S.K. Yang,
Nucl. Phys. \underbar{B315} (1987) 193.}

\REF\bd {T. Banks and L. Dixon, Nucl. Phys. \underbar{B307} (1988) 93.}

\REF\bdr {E. Bergshoeff and M. de Roo, Nucl. Phys. \underbar{B328}
(1989) 439.}

\REF\ns {D. Nemeschansky and A. Sen, Phys. Lett. \underbar{B178} (1986)
365.}

\REF\fmr {A.P. Foakes, N. Mohammedi and D.A. Ross, Nucl. Phys.
\underbar{B310} (1988) 335.}

\REF\z{B. Zumino, Phys.Lett.\underbar{B87} (1979) 205.}

\REF\ghr {S.J. Gates, C.M. Hull and M.Ro\v cek, Nucl. Phys.
\underbar{B248} (1984) 157.}

\REF\cmhb {C.M. Hull, {\sl Lectures on non-linear sigma models
and strings}, Super Field Theories Workshop, Vancouver 1986.}

\REF\dks {F. Delduc, S. Kalitsin and E. Sokatchev, CERN-TH-5521/89,
(1989).}

\REF\gsw {M.B. Green, J.H. Schwarz and P.C. West, Nucl. Phys.
\underbar{B254} (1985) 327.}
\REF\bds {A. Blasi, F. Delduc and S.P. Sorella, Nucl. Phys. \underbar
{B314} (1989) 409.}

\REF\wz  {B. Zumino, in
{\sl ``Relativity, Groups and Topology II''}, Ed. B.S. DeWitt and
R. Stora (1984) Amsterdam, North-Holland.}

\REF\ht {C.M. Hull and P.K. Townsend, Phys. Lett. \underbar{B178} (1986)
187.}

\REF\sen {A. Sen, Nucl. Phys. \underbar{B278} (1986) 289.}

\REF\hpc {P.S. Howe and G. Papadopoulos, Class. Quantum
Grav. \underbar {4} (1987), 1749.}

\REF\hw {C.M. Hull and E. Witten, Phys. Lett. \underbar{160B} (1985)
398.}

\REF\cfmp{C.G. Callan, D.Friedan, E.J. Martinec and M.J. Perry,
Nucl. Phys. \underbar{B262} (1985) 593.}

\REF\r {D.A. Ross, Nucl. Phys. \underbar{B286} (1987) 93.}

\REF\mt {R.R. Metsaev and A.A. Tseytlin, Phys. Lett. \underbar{B185}
(1986) 52.}

\REF\efs {U. Ellwanger, J. Fuchs and M.G. Schmidt, Nucl. Phys.
\underbar{B314} (1989) 175.}

%%%%%%%%%%%%%%%%%%%%%%%%%%%%%%%%%%%%%%%%%%%%%%%%%%%%%%%%%%%%%%%%%%%%%%%%%
\chapter{Introduction}

It has been known for some time that there several examples of
supersymmetric quantum field theories
which are finite to all orders in perturbation
theory (for a review see, e.g. [\hks]). Included in the list of finite
theories are some two-dimensional non-linear sigma models; in
particular, it was shown by Alvarez-Gaum\'e and Freedman [\agf] that
supersymmetric models with hyperk\"ahler target spaces have this
property (see also [\cmha,\gios]).
These sigma models have $N=4$ supersymmetry, or in chiral
notation, (4,4) supersymmetry. Using (conventional)
extended superfield power-counting
arguments, the authors of refs. [\hpa,\hpb] argued that sigma models
with (4,p) supersymmetry, $p=0,\dots 4$, are also be finite.
In the case of (4,0) supersymmetry this result was confirmed by
harmonic superspace methods in ref. [\kss] and by an analysis of the
(4,0) Ward identities in ref. [\bp].

Two-dimensional
supersymmetric sigma models with four-dimensional target spaces are of
interest in the context of solutions to heterotic
string theory for which spacetime has the form of a product of
six-dimensional Minkowski space and a non-trivial four-dimensional
Riemannian space. This type of solution is identified as a ``five-brane''
soliton in string theory and has been studied in a series of papers
by Callan, Harvey and Strominger [\s,\hs,\chsa,\chsb,\chsc].
(For further results on this subject see refs. [\dl,\pt].)
If the solution preserves six-dimensional
spacetime supersymmetry, then the sigma model whose target space is
the four-dimensional space transverse  to the five-brane
should  be (4,0) supersymmetric [\eoty,\bd,\chsa].
In references [\s,\hs,\chsa,\chsb,\chsc] several
explicit solutions have been studied, with particular emphasis being
placed on those which have (4,4) worldsheet supersymmetry.

In order to get a solution of string theory it is necessary that the
sigma model be superconformally invariant, a requirement
automatically met by finite sigma models. As we have remarked (4,4)
sigma models have this property [\agf], including
those with torsion [\hpa]. According to the results quoted above,
sigma models with (4,0) supersymmetry are also finite, but
the authors of refs.
[\chsa,\chsc] have
claimed that this is not the case, so that there appears to
be a contradiction. It is the purpose of this article to clarify the
issues that have arisen and to resolve this contradiction. We shall
confirm the finiteness of (4,0) sigma models explicitly up to three loop
order in perturbation theory and identify the corrections to the metric
as arising from finite local counterterms which must be added to the
theory in order to maintain (4,0) supersymmetry.

As we understand it, the argument that (4,0) sigma models are not finite
given in refs [\chsa,\chsc]
is based on the fact that the spacetime supersymmetry
transformations, computed for example in ref. [\bdr] in the effective
ten-dimensional supergravity theory for the massless modes of the string,
do not automatically vanish for background spacetimes corresponding to
(4,0) sigma models. This, it is claimed, shows that there must be higher
order corrections to the sigma model beta-functions. However, the
world-sheet sigma model for the heterotic string is naturally only (1,0)
supersymmetric; if a solution is (4,0) supersymmetric at lowest order in
$\alpha'$ it does not follow that such a sigma model remains (4,0)
supersymmetric at higher orders in perturbation theory. In fact, in
general, such a sigma model does not remain (4,0) supersymmetric unless
finite local counterterms are added. The introduction of the correct
finite corrections at $L-1$ loop order should, and indeed does, ensure
that the beta-functions vanish at the $L^{th}$ loop order. Thus, for
(4,0) theories there
are higher-order corrections to the metric, as the authors of refs.
[\chsa,\chsc] have stated, and when taken into account they
correct the spacetime supersymmetry transformations,
whereas this is not necessary in the (4,4)
case. However, these corrections come from finite local counterterms
and are moreover computable in terms of the tree level background fields.
{}From refs. [\hpa,\hpb,\kss,\bp]  we know that (4,0) superconformal
invariance can be implemented at all orders, so that (4,0) sigma models
do provide solutions to heterotic string theory of the type we have been
discussing, and the explicit forms of the background fields can be found
in principle by implementing (4,0) supersymmetry. This is somewhat
different to the case of $N=2$ sigma models on Calabi-Yau spaces, where
the corrections to the metric that are required to restore superconformal
invariance invole solving differential equations of progressively higher
order on the target space [\ns].

The organisation of the paper is as follows: in the next section we
briefly review (4,0) sigma models, in particular those for which the
target space is four-dimensional; in section 3 we analyse the one-loop
anomaly and show how it is cancelled; in section 4 we explicitly
compute the metric beta-function up to three-loop order using the results
of ref. [\fmr], and show that it vanishes to this order modulo sigma
model field redefinitions (diffeomorphisms of the target space). We
make some concluding remarks in section 5.
%%%%%%%%%%%%%%%%%%%%%%%%%%%%%%%% CHAPTER 2%%%%%%%%%%%%%%%%%%%%%%%%%%%%%%%%%%%%

\chapter{(4,0) sigma models}

The study of the interplay between the geometry of the target space
and extended supersymmetry in two-dimensional
supersymmetric sigma models began
with the realisation that $N=2$ supersymmetry requires the target space
to be a K\"ahler manifold [\z], and was followed by the result
that $N=4$ requires hyperk\"ahler target manifolds [\agf].
Models with torsion
and heterotic supersymmetry involve different geometries, and these
have been discussed by many authors; see, for example, refs.
[\ghr,\cmhb,\hpa,\hpb,\dks]. We shall follow [\hpa,\hpb]
here.

The classical action for the
$(1,0)$ sigma model is  $$ S[g,b,\phi]= -i \int d\sp{2}x d\theta^+\,
(\gij+\bij) D_+\ffi \del_= \ffj
\eqn \onea $$
where
$\phi$ is a map from $(1,0)$ superspace, $\Sigma\sp{(1,0)}$, with real
(light-cone)
co-ordinates $(x^{\pp}, x\sp{=}$, $\theta\sp{+})$, to the target space
$M$ given in local co-ordinates by $\phi^i(x^\pp,x^=,\th^+)$ ,$
i=1,\cdots,\ {\rm dim}\ M$. $M$ is  equipped
with a metric $g$ and a locally defined two-form $b$. The fermionic
derivative $D_+$ satisfies
$$ D_+\sp{2}=i\del_{\pp}.
\eqn\twoa$$

The action \onea\ is $(4,0)$ supersymmetric if $M$ admits three complex
structures $I_r$ $(r=1,2,3)$ obeying the algebra of the
imaginary unit quaternions,
$$ I_r I_s =-\delta_{rs} + \epsilon_{rst} I_t,      \eqn\oneb$$
if $g$ is Hermitian with respect to all three complex structures, and if each
of these is covariantly constant with respect to $\Gamma\sp{(+)}$:
 $$  g_{ij}I\sp{i}_{(r)k}I\sp{j}_{(r)l} = g_{kl}
\eqn\threeb$$
$$ \nabla\sp{(+)}_k I\sp{i}_{rj}= 0,
\eqn\twob
$$
where repeated indices in parentheses are not summed over and
$$\Gamma\sp{(\pm)i}_{jk}=\Gamma\sp{i}_{jk}\pm
{1\over 2} H\sp{i}_{jk}; \quad H_{ijk}=3 \del_{[i}b_{jk].}
\eqn\fivea$$

If $M$ is n-dimensional, then $n=4m$ for some integer $m$.  The additional
supersymmetry transformations are given by
$$\delta_r\ffi=i\zeta_{(r)} I\sp{i}_{(r)j} D_+\ffj
\eqn\fourb$$
where $\zeta_{r},r=1,2,3$ are the anticommuting parameters of the
extended supersymmetry transformations
which we take to be
constant.
If $E$ is a real rank q vector bundle over $M$ with connection $A$
and $\psi$ a
section of $\phi\sp{\ast}E\otimes S_-$ where $S_-$ is a
bundle of right-handed
spinors over $\Sigma\sp{(1,0)}$, then we can generalise the
$(1,0)$ action
\oneb\ by adding to it a fermionic sector with the action
$$S_{F}= -\int d\sp{2}x d\theta^+\, \psi\sp{A}_-
\nabla_+\psi\sp{A}_-;
\eqn\sevenb$$
where
$$ \nabla_+\psi\sp{A}_-=D_+\psi\sp{A}_-+
D_+\ffi A\sp{A}_{iB} \psi\sp{B}_-;\qquad A=1,\dots q.
\eqn\eightb$$ This is not the most general action of this type, but will
suffice for the present purpose.

The action \sevenb \ is $(4,0)$ supersymmetric under
$$\delta_r\psi\sp{A}_-= -i
\zeta_{(r)} I\sp{i}_{(r)j} D_+\ffj A\sp{A}_{iB}
\psi\sp{B}_- +  i\zeta_{(r)} \hat{I}\sp{A}_{(r)B}
\nabla_+\psi^B_-
\eqn\eightb$$
if the set of three complex structures $\hat{I}_r$ obeys the
algebra of the imaginary unit quaternions and if all three
are covariantly constant with
respect to $A$.  In addition, the curvature of $A$ must be $(1,1)$ with
respect to all three complex structures,
$$F_{ij}=F_{kl} I\sp{k}_{(r)i} I\sp{l}_{(r)j}
\eqn\nineb$$
The covariant constancy of the $\hat{I}$'s implies that the holonomy of the
connection $A$ must be a subgroup of $Sp(q/4)$,
$$\hat{I}\sp{A}_{rC} F\sp{C}_{ijB} = F\sp{A}_{ijC} \hat{I}\sp{C}_{rB}.
\eqn\tenb$$

If $M$ is a four-dimensional manifold then the equation \nineb\ is familiar
from
instanton physics; it is the condition for $F$ to be self-dual.  In addition it
was shown in reference [\chsa]
that the metric $g$ is conformally equivalent to
a Ricci flat metric $g\sp{o}$, i.e.
$$ g=\exp \Omega\ g\sp{o}
\eqn\tenab
$$
where $\Omega$ is harmonic with respect to the Laplacian of
the metric $g$ and $H=*d\Omega$, where $*$ denotes the Hodge
dual.
If $M$ is a compact manifold
without boundary, $\Omega$ is constant and the torsion $H$ vanishes.
In this paper, we shall be concerned with two classes of models:
for type one we will take the target space $M$ to be compact without
boundary , and hence
hyperk\"ahler (i.e. a $K3$ surface),
with a self-dual $Sp(q/4)$ gauge field, chosen so
that the instanton number is equal to the Euler number of $M$ ($K3$
has been studied in the context of string compactifications to
six dimensions previously [\gsw]); for
type two we shall consider instantons on $R^4$. We shall be able to
discuss the two cases together, since for both of them the Ricci tensor
and the torsion vanish at the zeroth order in $\alpha'$.

Since the algebra of supersymmetry transformations \fourb\ and \eightb\ closes
without the use of the field equations, we have an off-shell representation of
$(4,0)$ supersymmetry [\hpa, \hpb]. It is therefore possible to write
(4,0) sigma models in terms of (4,0) superfields, and thus to apply
extended superfield non-renormalisation theorems to them. In [\hpa,
\hpb] it was shown that this leads one to conclude that these models
should be finite. At first sight this might seem to be surprising as
(2,2) superspace has the same number of $\th$'s as (4,0) superspace
but (2,2) sigma models are not finite.
The crucial difference between the two cases is that the (4,0)
superspace measure is not two-dimensionally Lorentz invariant, so that
any counterterm Lagrangian will also not be a Lorentz scalar. For
a counterterm to be constructed from scalar fields, for example, it
follows that there must be derivatives in the counterterm Lagrangian,
and this immediately gives improved power-counting behaviour compared
to the (2,2) case. This property of the measure was also used in the
harmonic superspace analysis of the ultra-violet behaviour of (4,0)
sigma models [\kss].

%%%%%%%%%%%%%%%%%%%%%%%%%%%%%%%%% CHAPTER 3 %%%%%%%%%%%%%%%%%%%%%%%%%%%%%%%%%%%

\chapter{Anomalies}

It is convenient when discussing the anomalies to introduce a vielbein
$e\sp{a}_i$ on $M$; the spin connection
for the local orthonormal frame rotations which
corresponds to the Levi-Civita connection is
$\omega\sp{a}_{ib}$ and we define, as usual,
$$ \omega_i \sp{(\pm) ab} = \omega_i \sp{ab} \mp {1\over 2} H_i{}\sp{ab}.
\eqn\onec$$
The action \onea\ can be regarded as a function of $\phi$, $e$ and
$b$.  It is invariant under the (generalised) transformations
$$ \delta_l e\sp{a}_i = l\sp{a}_b(\phi) e\sp{b}_i, \  \delta_lb_{ij}=0, \quad
\delta_l\ffi=0
\eqn\twoc$$
$$\eqalign {
\delta_v e\sp{a}_i &=  v\sp{k} \del_k e\sp{a}_i +
\del_i v\sp{k} e\sp{a}_k, \
\delta_v\ffi=-v\sp{i}
   \cr
\delta_v b_{ij} &= v\sp{k} \del_k b_{ij} + \del_i v\sp{k} b_{kj}-\del_j
v\sp{k} b_{ki}          \cr}
\eqn\threec$$
$$ \delta_m b_{ij}= \del_i m_j- \del_j m_i      \eqn\fourc$$
where the parameters $l_{ab}(\phi)$ ($-l_{ba}\phi)$), $v\sp{i}(\phi)$ and
$m_i(\phi)$ correspond to local frame rotations, diffeomorhisms and
antisymmetric gauge transformations respectively.
For the fermion sector we have the gauge transformations
$$\eqalign {&
\delta_u \psi\sp{A}_- = u\sp{A}_B \psi\sp{B}_-, \quad \delta_u \ffi=0
\cr &
\delta_u A\sp{A}_{iB} = -\del_iu\sp{A}_B + u\sp{A}_C A\sp{C}_{iB} -
A\sp{A}_{iC} u\sp{C}_B  \cr}
\eqn\sixc$$
The above transformations are not symmetries in the usual field-theoretic
sense since they involve transforming the parameters (couplings)
(as well as the sigma model field in the case of diffeomorphisms).
Nevertheless, they lead to identities which are satisfied by the
classical action and which one wishes to preserve at the quantum level
in order to maintain the geometric interpretation of the parameters
as fields on the target space.
For the (4,0) models we also have the additional supersymmetry
transformations \fourb\ and \eightb\
which commute with the gauge transformations.

Sigma models are most easily quantised using the background field method.  The
quantum field $\xi\sp{a}(x,\theta)$ is a section of $\phi\sp{\ast}TM$ where
$\phi$ is the background field.  The background-quantum split introduces a new
shift symmetry, but a regularization scheme which preserves this symmetry and
the anti-symmetric tensor gauge invariance can always be found. A Ward
identity proof of the non-anomalous nature of the shift symmetry has
also been given [\bds].
Furthermore,
by a suitable choice of counterterms the theory can be made invariant under
diffeomorhisms.  Thus
we are led to consider the one-loop background effective
action $\Gamma[e,b,\phi]$ as a function of the parameters $e$, $b$ and the
background field $\phi$ which satisfies
$$\delta_v\Gamma=\delta_m\Gamma =0
\eqn\eightc$$
and
$$\delta_l\Gamma = \Delta (l)
\eqn\ninec$$
$$\delta_{\zeta_r}\Gamma = \Delta (\zeta_r)
\eqn\tenc$$
where $\Delta (l)$ and $\Delta (\zeta_r)$ are the potential obstructions;
they
are integrated local functionals of the background field $\phi$.  The
 Wess-Zumino
consistency conditions [\wz] include
$$ \delta_{l_1} \Delta (l_2) - \delta_{l_2} \Delta (l_1) = \Delta ([l_1,l_2])
\eqn\elevenc$$
$$ \delta_{l} \Delta (\zeta_r)  - \delta_{\zeta_r} \Delta (l) =0
\eqn\twelvec$$
$$ \delta_{\zeta_r} \Delta (\zeta_s) - \delta_{\zeta_s}
\Delta (\zeta_r) = 0.
\eqn\thirteenc$$
Equation \elevenc\ is the usual Wess-Zumino equation for the
gauge anomaly; its solution is
$$ \Delta (l) = -i k \int d\sp{2}x d\theta^+\,
Q\sp{1}_{ij}(l,\omega) D_+\ffi
\del_=\ffj
\eqn\fourteenc$$
where $\omega$ is an $SO(n)$ connection which can be arranged to be equal to
$\omega\sp{(-)}$ [\ht]. The constant
$k= {\ap\over 2}$ by an explicit calculation [\ht]. (In terms of
two-dimensional quantum field theory, $\ap={\hbar\over2\pi}$.)
 $Q\sp{1}_2$ is
 the usual two-dimensional anomaly occurring
 in the descent sequence [\wz]
$$P_4 \rightarrow Q\sp{0}_3 \rightarrow Q\sp{1}_2 \rightarrow Q\sp{2}_1
\rightarrow Q\sp{3}_0; \quad P_4 = {\rm Tr}
R\sp{2}(\omega).
\eqn\fifteenc$$
 \twelvec\ then implies that
$$ \Delta (\zeta_r) = 3 i k \int d\sp{2}x d\theta^+\,
\delta_r\phi^i
Q\sp{0}_{ijk}(\omega\sp{(-)}) D_+\ffj \del_=\phi\sp{k}  +
(l-{\rm invariant})
\eqn\sixteenc$$ The identities \thirteenc\ for $r=s$
will be satisfied by \sixteenc\ with no $l$-invariant term because
$P_4(\omega\sp{(-)})$ is a $(2,2)$ form with respect to all three
complex structures. The
cross-identities, $r\not= s$, of \thirteenc\ are also satisfied.
The absence of
$l$-invariant terms in \sixteenc\ can be confirmed by an explicit one loop
calculation.

        It is straightforward to include the fermion sector.
For the gauge anomaly,
the only changes are that $l$ is replaced by $u$ and $\omega$ by
$A$;  there is also a sign change for $\Delta(u)$ as compared to
$\Delta(l)$.  Furthermore, since $P_4(A)$ is $(2,2)$ with respect to
all three complex structures
the consistent (potential) supersymmetry anomaly is given by
\sixteenc\ with $Q^0_3(\omega^{(-)})$ replaced by $Q^0_3(\omega^{(-)})-
Q^0_3(A)$.

Having identified the potential obstructions to
implementing the symmetries
\twoc\ $\rightarrow$ \sixc\ quantum mechanically,
we need to examine whether or
not they can be removed by finite local counterterms,
i.e. we need to find a finite local
counterterm whose variation under supersymmetry gives precisely \sixteenc.

Since $P_4$ is a $(2,2)$ form it can be written in three different ways as
$$P_4=d d_{(r)} Y_{(r)}
\eqn\oned  $$
where $Y_r$ is a $(1,1)$ form with respect to $I_r$.  The derivative $d_r$
is defined as follows: for each complex structure we can define
holomorphic and anti-holomorphic derivatives $\partial_r$ and
$\bar\partial_r$; the exterior derivative $d$ is then the sum of the
two whereas $d_r$ is $i(\partial_r-\bar\partial_r)$. The Chern-Simons
three-form $Q^0_3$ can also be written in three different ways as
$$
Q^0_3 = dX_r + d_{(r)} Y_{(r)}
\eqn\threed$$
where the $X_r$ are two-forms.
The necessary and sufficient conditions for the
cancellation of supersymmetry
anomalies is the existence of a symmetric
second rank tensor $T$ which is hermitian with
respect to all the
complex structures and a two-form (possibly locally defined on
$M$) $X$ such that
$$ Y_{rij} = T_{ik} I\sp{k}_{rj}\qquad {\rm and}\qquad X=X_r
\eqn\fourd$$
Indeed, in this case there is a finite local counterterm for which the
action, $S_c$, is
$$S_c =- {i\ap\over
2}  \int d\sp{2}x d\theta^+\, (T + X)_{ij} D_+ \ffi \del_=\ffj
\eqn\fived $$
This finite local counterterm, if it exists, cancels the extended
supersymmetry anomalies and the
$SO(n)$ anomaly as well.  This can be shown by a
straightforward computation, and is similar to the anomaly cancellation
process in $(2,0)$ models [\sen,\hpc]. The $SO(n)$ transformation
property of $X$ is
$$
\delta_l X= -Q^1_2(l,\omega^{(-)})
\eqn\fiveaad
$$
so that if $b$ is redefined by $b\rightarrow\bar b=b+{\ap\over2}X$,
the field $\bar b$ has the familiar ``anomalous'' transformation
rule $\delta \bar b=-{\ap\over2} Q^1_2(l,\omega^{(-)})$
[\hw] (and similarly
for gauge transformations).
There may be some remaining symmetries in
the theory like the holomorphic symmetries for the $(2,0)$ case [\hpc].
However, these do not cause any problems; we shall comment on them a
little later.
Finally, there is a global reparametrisation
anomaly in the theory which is
cancelled provided that $P_4$ is an exact form on $M$.

The analysis given so far is applicable to any (4,0) target space.
We shall now focus on the case of four-dimensional target spaces where
there are some simplifications.
For both types of model we shall consider we can write $P\equiv
P_4(\omega)-P_4(A)$ in the form
$$
P= d*d f
\eqn\fivead $$
where $f$ is a differentiable
function on the target space. This is not difficult
to see in the compact case, using the Hodge decomposition, and may be
explicitly verified in the instanton case. For this latter case $f$ is
given by
$$
f=\triangle {\rm Tr}\log h
\eqn\fivebd$$
where $\triangle$ is the ordinary Laplacian on $R^4$.
The matrix $h$, which is a $k\times k$ matrix for an instanton
with instanton number $k$, is given by
$$
h\sp{-1}=  a_{ij} y\sp{i} y\sp{j}- b_i y\sp{i}+c
\eqn\sevend
$$
where $a$, $b$ and $c$ are $k\times k$ matrices
with entries which are the instanton parameters, and $y^i$ are Cartesian
co-ordinates on $R^4$.
In both cases the
symmetric tensor  $T$ is given by
$$ T_{ij} = -3 f g_{ij}
\eqn\eightd$$
and is clearly  hermitian with respect to all three
complex structures. We can also show that there is an antisymmetric
tensor $X$ that satisfies \fourd. To prove this observe that for
four-dimensional manifolds the second term in the right-hand-side
of \threed\ can be written in such a way that the
complex structures do not
appear explicitely by using the $\epsilon -$tensor.
Hence the supersymmetry and
gauge anomalies are cancelled in this case. The
remaining symmetry of the theory is the ambiguity
in specifying the instanton
parameters of a given self-dual connection. This is a rigid
symmetry and
the finite local counterterm is invariant under it.

%%%%%%%%%%%%%%%%%%%%%%%%%CHAPTER 4%%%%%%%%%%%%%%%%%%%%%%%%%%%%%%%%

\chapter{The beta function}

We now turn to the evaluation of the metric $\beta$-function.
Our strategy is as follows: the (1,0) sigma model $\beta$-function
for the metric
has been computed up to two [\cfmp,\cmhb]
and three loops [\fmr]. We shall
express
this $\beta$-function in terms of parameters $\bar g,\bar b$ and then
correct it order by order in $\ap$ by demanding (4,0) supersymmetry.
Thus at lowest order, $\bar g=g$ and $\bar b=b=0$, where $g$ is the
(4,0) metric, while
at first order, we must add the finite local counterterm \fived\ in
order to cancel the anomalies. This amounts to a redefinition of the
parameters by
$$
\bar g_{ij} = g_{ij} - {3\over2}\ap f g_{ij}
\eqn\threee
$$
and
$$
\bar b_{ij} = b_{ij} + {\ap\over2} X_{ij}
\eqn\foure
$$
We then substitute these expressions into the two-loop $\beta$-function
and evaluate it to order $\app^2$ in terms of the parameters $g,b$.
Having done this, we go to the next order which will entail adding
$\app^2$ terms to the right-hand sides of \threee\ and \foure.
In principle, we could calculate these second-order terms by looking
at the two-loop anomaly, but we have not done this. However, we know
they must exist, and it is easy to guess the correct form for the
metric from the three-loop $\beta$-function. Because $b=0$, the
second order correction to $b$ is not needed for the order to which
we are working. In this way we are able to calculate the $\beta$-function
(for $\bar g$) order by order in $\ap$ in terms of the ``known'' metric
$g$ (as well as the gauge field $A$). Not surprisingly, it vanishes up
to a Lie derivative of the metric
which can be removed by a sigma model field
renormalisation. When this last step has been carried out, the
$\beta$-function for $\bar g$ is zero (to this order) and so the
$\beta$-function for $g$ is also zero. By the general results of refs
[\hpa,\hpb,\kss,\bp] we know this can be done to all orders, which
means that there is a renormalisation scheme, using the parameters
$g,b,A$, in which the metric $\beta$-function vanishes order by
order in perturbation theory, i.e. the theory is finite.

We now turn to the details of the three-loop calculation outlined above.
Up to two loops the (1,0) sigma model metric $\beta$-function is [\cmhb]
$$
\beta_{ij}(\bar g) = -\alpha' R^{(+)}_{(ij)} -{(\alpha')
^2\over4} {\rm Tr}
\big(R^{(-)}_{ik} R^{(-)k}_j - F_{ik} F_j{}^k\big)
\eqn\onee
$$
expressed in terms of parameters $\bar g,\bar b$ (and $\bar A$, which
we can set equal to $A$).
The
curvature tensors $R^{(+)}$ and $R^{(-)}$ are computed from the
connections $\Gamma^{(+)}$ and $\Gamma^{(-)}$ repectively. They obey
the identity
$$
R^{(+)}_{ij,kl} = R^{(-)}_{kl,ij}
\eqn\twoe
$$
where the first pair of indices are the Lie algebra indices, and the
second pair the differential form indices. The trace in \onee\
 is over
the Lie algebra indices. Because of \twoe, the Ricci tensors are not
symmetric and this necessitates the symmetrisation in the one-loop
contribution. Alternatively, we can write
$$
R^{(+)}_{(ij)} = R_{ij} -{1\over4}\tilde H_i{}^{kl}\tilde H_{jkl}
\eqn\twoae
$$
where $\tilde H$ is the Chern-Simons corrected torsion, i.e
$$
\tilde H = d\bar b - {\ap\over2} (Q^0_3(\omega^{(-)})-Q^0_3(A))
\eqn\twobe
$$
The $(+)$ and $(-)$ connections appearing in equations \onee\-\twobe\
above now involve the corrected torsion
$\tilde H$, and so must be found iteratively. It is usually assumed
that the Chern-Simons terms appear in this way, and this has
not been explicitly verified except at two loops [\r];
for our purposes, though,
it will only be necessary to consider the Chern-Simons term for the
standard spin connection of the metric $g$.
In fact, because $b$ is zero, the $H^2$ term does not play any role
until $O\app^3$. Following the strategy outlined above, it is clear
that the $O\app$ contribution to the beta-function vanishes as $g$
is Ricci-flat. To the next order, we have to expand $R_{ij}(\bar g)$
to order $\ap$
and include the lowest order contribution from the two-loop term.
Using the easily proved relation,
$$
{\rm Tr} \big(R_{ik} R_j{}^k - F_{ik} F_j{}^k\big)=
{1\over4}g_{ij} {\rm Tr}\big(R^2-F^2\big)
\eqn\fivee
$$
valid for self-dual curvature tensors, one can show
that the $\beta$-function reduces to
$$
\beta_{ij}(\bar g) = -{3\over2}
(\alpha')^2 \nabla_i\nabla_j f
\eqn\sixe
$$
to this order.
This remaining $(\alpha')^2$ contribution is thus
of the form $\nabla_{(i}
v_{j)}$ and so can be removed by making a suitable redefinition of
the sigma model field. Hence the $\beta$-function vanishes to two-loop
order.

We now turn to the three-loop calculation. This was carried out by
Foakes et al [\fmr], and their result agrees with the calculation
carried out from a string theory point of view by Metsaev and Tseytlin
[\mt]. The complete expression for a general (1,0) model is quite
complicated, but for our purposes we need only evaluate it for the
metric $g$, with $b$ set equal to zero. In this case nearly all the
terms drop out due to Ricci-flatness, and we are left with
the contribution
$$
\beta^{(3)}_{ij} =-{ (\alpha')^3\over32} \triangle
{\rm Tr} \big(R_{ik} R_j{}^k- F_{ik} F_j{}^k\big)
\eqn\sevene
$$
Due to the identity \fivee, this can be written, to the desired degree
of accuracy, as
$$
\beta^{(3)}_{ij} ={3(\alpha')^3\over32} \triangle^2 f g_{ij}.
\eqn\eighte
$$
Clearly, this term can be removed by a two-loop adjustment to the metric
which should be taken to be $-{3\over16}(\alpha')^2 \triangle\ fg_{ij}$.
Notice again that
it is a finite adjustment at the two-loop level which produces the three
loop cancellation in the $\beta$-function. However, we are not finished
yet as we should also take into account the effect of the one-loop
counterterms at three loops.

There are three contributions: the first is from the $O\app^2$ terms
in $R_{ij}(\bar g)$, the
second from the $H^2$ term which is contained in $R^{(+)}_{(ij)}$, and
which contributes first at three-loop order, and the third is from
making the first-order shift in the two-loop $\beta$-function, not
forgetting the term \sixe. When all these contributions are summed
the complete result, to three-loop order, is
$$
\beta_{ij}(\bar g) = \bar\nabla_{(i}v_{j)}
\eqn\ninee
$$
where
$$
v_i = \bar\nabla_i\big(-{3\app^2\over2}f-{9\app^3\over8}f^2
-{3\over16}\app^3\triangle f\big)
\eqn\tene
$$
Thus the three loop result reduces to a Lie derivative which can be
removed by a wave-function renormalisation. Hence we have explicitly
verified the superconformal invariance of the model up to three loops
and have shown that the finite counterterms required to restore (4,0)
supersymmetry are responsible for the finiteness of the theory, at
least in the metric sector of the theory. The expression for the
metric $\bar g$ is, to this order
$$
\bar g_{ij} = g_{ij} -{3\over2}\ap f\,g_{ij} -{3\over16}\app^2
\triangle f\,g_{ij}.
\eqn\elevene
$$

To our knowledge, there
does not exist a complete three-loop calculation of all of the
$\beta$-functions in a general (1,0) sigma model, but
a three-loop calculation
has been made of $\beta$-function for $A$ in the case of a background
gauge field only (i.e. a trivial metric) [\efs]. This result covers
the case of the gauge solution, and it is a short computation to
verify that the $\beta$-function for $A$ vanishes up to three loops
when $F$ is self-dual. Explicitly, the $\beta$-function for $A$ is
$$
\beta_i^{AB}=-{1\over4}\ap\nabla^j F^{AB}_{ij} -{3\over64}\app^3
\big[\nabla_k F^{AB}_{ij} F^{kCD}_{\ \,l} +2F^{kAB}_{\ \,j}
\nabla_{(k}F^{CD}_{l)i}\big] F^{jlCD}.
\eqn\twelvee
$$
The first term clearly vanishes when $F$ is self-dual, and it is not
difficult to see that the three-loop contribution does also. Further,
the conformal rescaling of the metric induced by the finite local
counterterms does not affect this result.

%%%%%%%%%%%%%%%%%%%CHAPTER 5%%%%%%%%%%%%%%%%%%%%%%%%%%%%%%%%%%%%%%%%%

\chapter{Conclusions}

In this article we have argued that there exists a renormalisation
scheme for the (4,0) sigma model in which the metric $\beta$-function
vanishes order by order in perturbation theory, and we have explicitly
confirmed this up to three loop order. This scheme is the natural one
from the point of view of (4,0) world-sheet supersymmetry, whereas
the scheme using $\bar g$ and $\bar b$ is more natural from the point
of view of (1,0) supersymmetry. The difference between two methods of
calculation in a field theory should only result in a change of
scheme as far as the $\beta$-functions are concerned; thus, although
it is more difficult in the (4,0) formalism to see explicitly what
is going on in terms of the metric, etc., since this method of
calculation manifestly preserves (4,0) supersymmetry,
it must be the case that
the (finite) counterterms which arise in the (1,0) computation are
automatically included in the (4,0) calculation. On the other hand,
from the point of view of string theory, it is the $\bar g$ metric
which is more natural, since the heterotic string is in general only
(1,0) supersymmetric. The $\beta$-function for this metric is not
zero order by order, as we have seen, although it does vanish when
summed. To this extent, the authors of refs. [\chsa,\chsc] have a
point, although we would maintain that it is the metric $g$ which
is natural in (4,0) supersymmetry.

The parameters $\bar g$ and $\bar b$
do not satisfy the (4,0) conditions given in
section 2, but instead satisfy more complicated constraints. It
would be of interest to determine these equations and to compare
them with spacetime results, and this is under investigation.
Finally, we have concentrated in this paper on the case of
four-dimensional target spaces as these are both topical and
simpler to deal with. However, we believe that these results
can be extended to the general case, where the dimension of
the target space is a multiple of four, and our belief is
fortified by the general results of refs. [\hpa,\hpb,\kss,\bp].

\vskip 1cm
\leftline{\bf Acknowledgement.}
We would like to thank K.S. Stelle for interesting discussions.
PH thanks CERN for its hospitality. GP thanks the SERC for financial
support.

\refout

\end